# Aging in a Colloidal Glass in Creep Flow: Time-Stress Superposition

Yogesh M Joshi* and G. Ranjith K. Reddy

Department of Chemical Engineering, Indian Institute of Technology Kanpur,
Kanpur 208016, INDIA.

* Corresponding Author, E-Mail: joshi@iitk.ac.in.

Tel: 0091 512 259 7993

Fax: 0091 512 259 0104

**Abstract**

In this work, we study ageing behavior of aqueous laponite suspension, a model soft glassy material, in creep. We observe that viscoelastic behavior is time dependent and is strongly influenced by the deformation field; the effect is known to arise due to ageing and rejuvenation. We show that irrespective of strength of deformation field (shear stress) and age, when imposed time-scale is normalized with dominating relaxation mode of the system, universal ageing behavior is obtained demonstrating time-stress superposition; the phenomena that may be generic in variety of soft materials.

~~~~~~~~~~~

Glasses are out of equilibrium materials and explore only a part of the phase space available to them. In molecular glasses rapid decrease in temperature leads to an ergodicity breaking while in colloidal glasses insufficient mobility due to crowding of arrested entities leads to ergodicity breaking.[1] Common in both the glasses is the ageing phenomenon that lowers the potential energy of the system with time.[2] In soft (colloidal) glasses ageing is achieved by activated dynamics of the individual arrested entities thereby lowering the energy.[3] Understanding dynamics of such 'out of equilibrium' materials, for example, pastes,[4] gels,[5] concentrated suspensions[6-9] and emulsions[10, 11] and other soft materials[12, 13] is important from both industrial as well as academic point of view. Prominent characteristic feature common in these materials is extremely slow relaxation behavior with dominant mode scaling with the age of the system. Clearly properties of these materials strongly depend on the deformation history, which leads to significant obstacles in analyzing and predicting behavior of these materials. Application of deformation field weakens dependences of dominant mode on age, the phenomenon generally addressed as rejuvenation.[4] In this paper we exploit this behavior. Since the rheological behavior is intrinsically dependent on characteristic timescale of the system, weakening dependence of the same on age at large stresses suggests a possibility of predicting rheological response at large stresses by carrying out short time tests at small stresses thereby leading to a "time-stress" superposition.

Most of the soft glassy systems rejuvenate under flow field, so that aging under flow is very slow compared to that under "no flow" conditions,[14] however in order to observe influence of flow, its timescale needs to be smaller than the dominating timescale of the material.[15] Thus, strength of flow field changes extent of rejuvenation, which may affect distribution of relaxation time differently.[6] Very recently Wyss et al.[16] demonstrated strain-rate frequency superposition and suggested that structural relaxation driven by imposed strain rate results in the same response as the equilibrium structural relaxation at much larger time scale (lower frequency). Cloitre et al.[4] observed that creep curves at different ages get superimposed when plotted against $t/t_w^\mu$, where $t$ is creep time, $t_w$ is age and $\mu$ is a positive exponent first introduced by Struik[17] for ageing amorphous polymers [also see Fielding et al.[18]]. They inferred that characteristic relaxation time of the system scales as age ($\tau \sim t_w^\mu$) with $\mu$ depending on the creep stress. Derec et al.[8, 19] carried out similar superposition for stress relaxation experiments for concentrated suspension of silica. Significantly, in this work, we take advantage of the weak dependence of characteristic relaxation time on age at large stress to predict rheological behavior at large stress by carrying out short time tests at small stresses. We observe that irrespective of value of creep stress (or $\mu$), system shows universal creep behavior by carrying out systematic shifting procedure thereby demonstrating time-stress superposition. Moreover we also observe that stress plays the same role in colloidal glasses that temperature plays in molecular glasses while reversing the effect of ageing.

Laponite RD, synthetic hectorite clay, is composed of disc shaped nanoparticles with a diameter 25 nm and a thickness 1 nm. Suspension of laponite in water shows two stage evolution of $\alpha$ relaxation mode and undergoes ergodicity breaking over practical time scales.[20] Laponite RD used in the present experiments is procured from Southern Clay Products, Inc. The white powder of Laponite was dried for 4 hours at 120 °C before mixing with deionized water at pH 10 under vigorous stirring for 15 min. The couette cell of stress controlled rheometer, AR 1000 (bob diameter 28 mm with gap 1mm) was filled up with the sample. After keeping the system idle for 3 hours, oscillatory deformation with stress amplitude of 50 Pa and frequency 0.1 Hz was applied for a short while to carry out shear melting. Sample yields under such a high stress and eventually shows a plateau of low viscosity that does not change with time. We stopped the shear melting experiment at this moment, from which the aging time ($t_w$) was measured. Shear melting procedure is used to achieve uniform initial state for all the samples used in the present study. Subsequently, in the ageing experiments, oscillatory shear stress with amplitude 0.5 Pa and frequency 0.1 Hz was employed to record the ageing behavior with respect to $t_w$. The detailed experimental procedure is also reported elsewhere.[9] All the results reported in this paper relate to 3.5 wt. % laponite suspension and temperature 20 °C. To avoid evaporation of water or $CO_2$ contamination of the sample, the free surface of the suspension was covered with a thin layer of low viscosity silicon oil throughout the experiment. We also carried out frequency sweep experiment on independent samples at the end of waiting period of 3 hours. We observed that storage modulus is independent of frequency while loss modulus showed slight decrease with respect to frequency in the experimentally accessible frequency range. This observation is similar to that observed by Bonn et al.[21] and according to Fielding et al.[18] this ensures system to be in the non-ergodic regime.

Figure 1 shows evolution of complex viscosity for a shear rejuvenated laponite suspension, wherein corresponding increase in complex viscosity is plotted against the age of the sample. We carried out aging experiments for all the independent samples until the predetermined value of age (corresponding complex viscosity between 300 Pas to 1400 Pas) is reached. After stopping the oscillatory test each time, creep experiments were performed with various creep stresses in the range 0.5 Pa to 5 Pa. As shown in an inset of fig. 1, in the initial period up to $O$ (1) s, the system shows damped oscillations in strain which are known to occur due to instrument inertia coupled with viscoelastic character of the fluid.[9]

As sample ages, its relaxation dynamics becomes slower and slower. We have discussed above that dominant relaxation time of the system scales as age ($\tau \sim t_w^\mu$) with $\mu$ depending on the creep stress. Thus normalization of creep time by $t_w^\mu$ is expected to give superposition of creep curves after appropriate vertical shifting of the creep curves is carried out. Vertical shifting is needed to accommodate thixotropic character of the system that is responsible for significant increase in modulus with time shown in fig. 1. In fig. 2 we plot creep curves obtained at stress of 3 Pa with ratio of creep compliance $J(t_w+t)$ and zero time compliance $J(t_w)\left[=1/G(t_w)\approx 1/G'\right]$ as ordinate and $t/t_w^\mu$ as abscissa. We have considered creep data beyond $t>1$ s only, to omit the oscillatory part. It can be seen that for $\mu=0.45$, the creep curves indeed superimpose on each other.

In order to understand this behavior lets assume that the system is represented by a Maxwell model (an elastic spring and a dashpot in series). In principle, in order to observe damped inertial oscillations, the spring of the Maxwell model needs to be replaced by a Kelvin-Voigt element (an elastic spring and a dashpot in parallel);[9] however since we are omitting the inertial oscillations, consideration of simple Maxwell model suffices the present purpose. Stain induced by application of step stress ($\sigma$) to the Maxwell model is then given by:

$$\gamma(t_w+t)=\frac{\sigma}{G}+\int_0^t \frac{\sigma}{\eta}dt. \qquad (1)$$

Here $G$ and $\eta$ represents modulus and viscosity associated with the Maxwell model. Viscosity of the Maxwell model can be represented as $\eta=G\tau$, where $\tau$ is the characteristic relaxation time of the system. Struik[17] (and more formally Feilding *et al.*[18]) suggested that $\tau$ scales as,

$$\tau = A\tau_0^{1-\mu} t_w^{\mu}, \tag{2}$$

where $A$ is a numerical factor, $\tau_0$ is a microscopic time[18] and $\mu$ is a function of stress. Incorporation of eq. (2) in eq. (1) gives:

$$\gamma(t_w + t) = \frac{\sigma}{G} + \frac{\sigma \tau_0^{\mu-1}}{GA} \int_0^t \frac{1}{(t_w + t)^{\mu}} dt. \tag{3}$$

We have taken $G$ out of integration assuming that it changes negligibly over the creep time in the limit of $t \ll t_w$. Integration in Eq. (3) can be easily solved and after rearrangement in the limit of $t \ll t_w$, we get:

$$J(t_w + t) G(t_w) = 1 + \frac{\tau_0^{\mu-1}}{A} \left( \frac{t}{t_w^{\mu}} \right). \tag{4}$$

Eq. (4) shows that after appropriately carrying out the vertical and horizontal shifting, as represented by a term on the left hand side and a second term of the right hand side respectively, a unique value of $\mu$ should show superposition.

By adopting same procedure as in fig. 2 we obtained superposition for other creep stresses as well. Figure 3 shows value of $\mu$ for which master curve is obtained plotted against creep stress. It can be seen that larger the value of stress, smaller is the value of $\mu$. As stress increases, it rejuvenates the system to a greater extent, reducing the effect of ageing. For creep stress of 5 Pa, complete rejuvenation occurs (no ageing) and we get $\mu = 0$, showing that yield stress of the present system is between 3 Pa to 5 Pa. Thus, $\mu$ depends on the extent of ageing and/or rejuvenation. When rejuvenation completely erases the effect of ageing, $\mu \to 0$; while for simple ageing, theoretical higher limit on $\mu$ is unity.[17, 18] Interestingly for polymeric glasses Struik[17] also observed decrease in $\mu$ with increasing stress; however now it has been accepted that mechanical rejuvenation may not be responsible for this effect.[22, 23] For structural glasses $\mu$ gets affected by temperature and strain. Struik[17] and more recently O'Connell and McKenna[24] observed that for polymer glasses, $\mu$ rapidly approaches zero as temperature approaches $T_g$. Thus, for molecular glasses, increase in temperature reverses the effect of ageing; while for soft glassy systems, application of stress reverses the effect of ageing.

We observed that when creep time is normalized by $t_w^\mu$ and compliance is normalized by $G(t_w)$, creep curves get superimposed for a particular value of $\mu$, which is a function of stress. In this study we employ four creep stresses leading to four sets of superimposed creep curves each having a unique value of $\mu$ as shown in fig. 3. Each of this set of superimposed creep curves can then be shifted horizontally by multiplying by a factor $a(\mu)$ to get a universal master curve. In fig. 4, 17 different creep curves at different stresses and ages (different values of $\mu$) are shown to get superimposed to form a universal master curve. We have shifted all the sets of creep curves on to a set of creep curves with $\mu$=0.45 (creep stress 3 Pa) by carrying out a horizontal shifting by varying parameter $a(\mu)$. The minor scatter in the universal master curve is due to decaying inertial oscillations that are observed beyond creep time $O$ (1s) as shown in the inset of fig. 1. Eq. 2 suggests that $a(\mu)$ should have a logarithmic dependence on $\mu-1$ given by $a \sim \tau_0^{\mu-1}$. Inset of fig. 4 shows a semi-log plot of $a$ vs. $\mu-1$. It can be seen that a straight line representing $a \sim \tau_0^{\mu-1}$ indeed fits the data very well. The same plot can be represented as a time-stress (or time-$\mu$) superposition with top abscissa corresponding to time axis. Such superposition represents all the creep curves with different values of $\mu$ (different creep stress) and different ages shifted on a creep curve with $\mu$=0.45 and $t_w$=25272 s. The shift factor $b$ can be represented by, $b = \tau(\mu_0, t_{w0})/\tau(\mu, t_w) = \tau_0^{\mu-\mu_0} t_{w0}^{\mu_0}/t_w^\mu$, where $t_{w0}$ and $\mu_0$ are the values pertaining to the reference creep curve (in present case $\mu_0$=0.45 and $t_{w0}$=25272 s).

A fit to estimated values of $a$ with respect to $\mu-1$ as shown in an inset of figure 4, leads to independent estimation of the product $A\tau_0$ to be equal to around 53 s. This is the characteristic time-scale of the system in the limit of $\mu \to 0$. In order to validate the same quantitatively, the shear viscosity and modulus of the samples wherein creep stress of 5 Pa ($\mu = 0$) was applied was used to estimate the characteristic time scale ($\tau \approx \eta/G$). We observed that the relaxation time computed from shear viscosity and modulus comes out to be in the range 60 s to 80 s for various samples, and is very close to a value 53 s obtained from the inset of figure 4. This observation clearly validates the present protocol quantitatively.

Cloitre et al.[4] showed that creep curves at different ages but same creep stress can be shifted to form a master curve. In the present manuscript we show that creep curves at different stresses and creep times can also be superimposed to form a universal master curve. Furthermore, the shift factor required for time superposition shown here is derived from expression for the dominating relaxation time (Eq. 2) proposed by Fielding et al.[18] We believe that time-stress superposition shown in fig. 4. is a significant result which can be attributed to sole dependence of rheological behavior on dominating relaxation time which is altered by stress and age. Deformation field or stress retards the structural evolution (or ageing) by influencing the activated dynamics. Consequently ageing that occurs for a shorter period of time at lower stresses is same as what occurs for a longer period of time at larger stresses. Therefore one can predict the later behavior from the short time test at low stress as shown in fig. 4.

In conclusion, we investigate ageing and rejuvenation of soft solids of aqueous suspension of laponite using creep experiments. We observe that characteristic relaxation time shows a power law dependence on age. The corresponding power law index characterizes extent of rejuvenation such that it decreases when higher stress rejuvenates the system. Significantly, this study shows that when creep time is normalized with dominating relaxation mode, which is dependent on age and creep stress, the universal master curve demonstrating time-stress superposition is obtained. We explicitly show that stress plays the same role in soft glassy materials as temperature plays in structural glasses while reversing the extent of ageing. We believe that various features of ageing in soft materials discussed in the present paper are generic and can be applied to variety of soft glassy materials.

Financial support from Department of Atomic Energy, Government of India under the BRNS young scientist award scheme is greatly acknowledged.

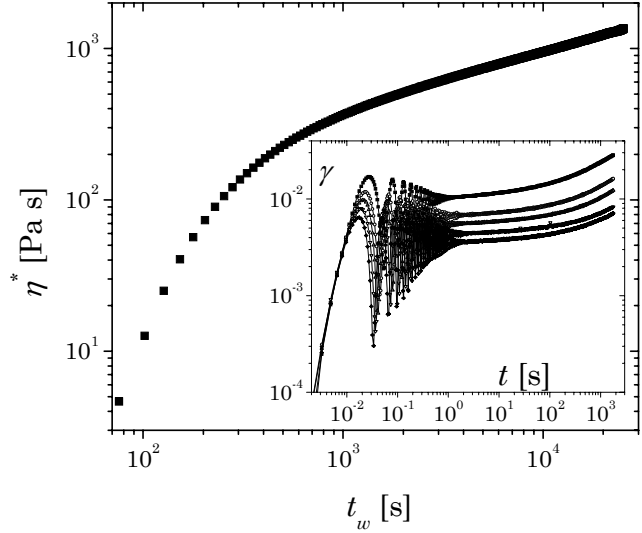

**Figure 1.** Ageing curve for 3.5 wt. % aqueous laponite suspension. Inset shows creep curves obtained at various ages (from top to bottom: 1908 s, 5430 s, 9719 s, 18725 s, 25272 s) for creep stress of 3 Pa.

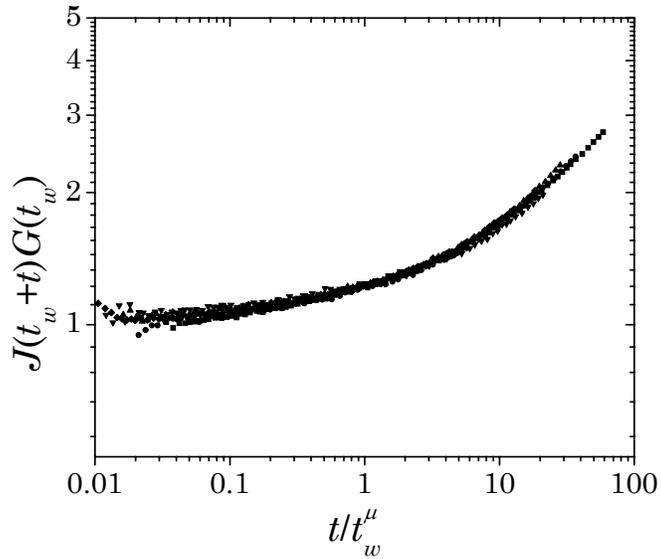

**Figure 2.** Normalized creep compliance plotted against $t/t_w^\mu$ for $\mu=0.45$ (creep stress=3 Pa). At this value of $\mu$, creep curves obtained at different ages show superposition. (squares 1908 s, circles 5430 s, up triangles 9719 s, down triangles 18725 s and diamonds 25272 s)

arXiv:0710.5264          8

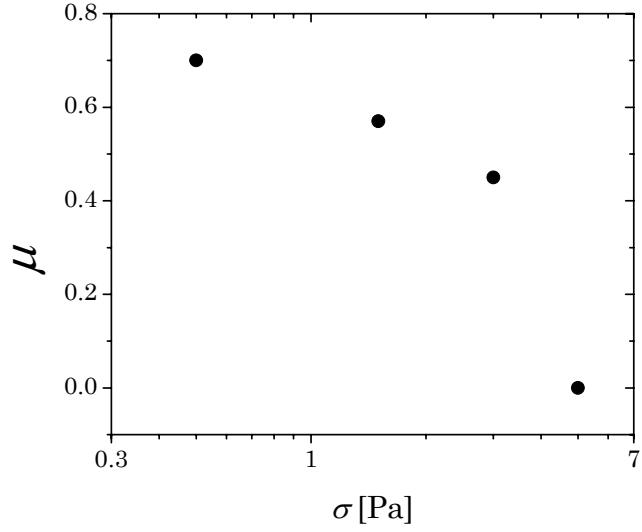

**Figure 3.** Parameter $\mu$ is plotted as a function of creep stress. Increase in stress enhances extent of rejuvenation causing decrease in $\mu$.

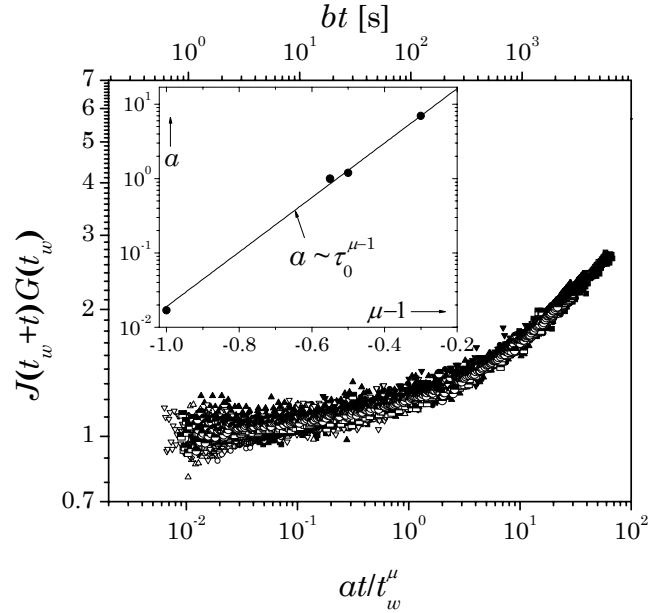

**Figure 4.** Universal master curve for different ages and creep stresses. Parameter $a$ is a shift factor for horizontal shifting and inset shows its dependence on $\mu-1$ follows: $a \sim \tau_0^{\mu-1}$. Shift factor $b$ is given by, $b = \tau_0^{\mu-\mu_0} t_{w0}^{\mu_0} / t_w^{\mu}$. (see text for discussion).